\documentclass[sigconf]{acmart}
\usepackage{booktabs} 
\usepackage{xspace}

\usepackage{multirow} 
\usepackage{array} 
\usepackage{tabularx} 
\usepackage{enumitem}

\usepackage{algorithm,algpseudocode}
\usepackage{amsmath}
\usepackage{amsfonts}

\usepackage{stmaryrd}

\graphicspath{{Figures/}}

\setlist{leftmargin=5.5mm}

\newcommand{\ie}{\emph{i.e.,}\xspace}
\newcommand{\eg}{\emph{e.g.,}\xspace}

\newcommand{\paratitle}[1]{\vspace{1ex}\noindent \textbf{#1}}

\newcommand{\mmatch}{M-Match\xspace}

\def\BibTeX{{\rm B\kern-.05em{\sc i\kern-.025em b}\kern-.08emT\kern-.1667em\lower.7ex\hbox{E}\kern-.125emX}}

\begin{document}

\title{Mirror Matching: Document Matching Approach in Seed-driven Document Ranking for Medical Systematic Reviews}

\author{Grace E. Lee}
\authornote{This work has been done when she was a PhD candidate.}
\affiliation{%
  \institution{School of Computer Science and Engineering, Nanyang Technological University, Singapore}
}
\email{leee0020@e.ntu.edu.sg}
\author{Aixin Sun}
\affiliation{%
  \institution{School of Computer Science and Engineering, Nanyang Technological University, Singapore}
}
\email{axsun@ntu.edu.sg}

\begin{abstract}
When medical researchers conduct a systematic review (SR), screening studies is the most time-consuming process: researchers read several thousands of medical literature and manually label them relevant or irrelevant. Screening prioritization (\ie document ranking) is an approach for assisting researchers by providing document rankings where relevant documents are  ranked higher than irrelevant ones. Seed-driven document ranking (SDR) uses a known relevant document (\ie seed) as a query and generates such rankings. Previous work on SDR seeks ways to identify different term weights in a query document and utilizes them in a retrieval model to compute ranking scores. Alternatively, we formulate the SDR task as finding similar documents to a query document and produce rankings based on similarity scores. We propose a document matching measure named Mirror Matching, which calculates matching scores between medical abstract texts by incorporating common writing patterns, such as background, method, result, and conclusion in order. We conduct experiments on CLEF 2019 eHealth Task 2 TAR dataset, and the empirical results show this simple approach achieves the higher performance than traditional and neural retrieval models on Average Precision and Precision-focused metrics.
\end{abstract}


\maketitle

\section{Introduction}
\label{sec:Introduction}

In evidence-based medicine, a systematic review (SR) plays an important role in informing healthcare professionals and policy makers. It provides a conclusive answer for a clinical question by synthesizing all existing studies. When researchers write a SR,  finding \textit{all} relevant studies is a critical goal to derive a reliable answer. Researchers strive for the goal by carefully following sequential steps: (i) define a clinical question and relevance conditions, (ii) retrieve documents (candidate documents) from digital libraries, (iii) screen the candidate documents to identify relevant documents, (iv) analyze and synthesize the relevant documents and derive an overall conclusion~\cite{SR_survey}.

Screening is the most time-consuming and expensive step above all because of exhaustive manual labeling. SR researchers read abstracts of all candidate documents (typically around 6,000 documents) and identify intermediate relevant documents based on the content presented in abstracts  (\ie abstract stage screening). They then prepare and read full documents of the intermediate relevant documents, and determine final relevant documents (\ie full document stage screening). This manual and multi-stage screening step takes weeks to months to complete.

Screening prioritization (\ie document ranking) is an approach for assisting researchers by providing document rankings where relevant documents are ranked higher than irrelevant documents. This approach makes the time-consuming screening step more efficient by allowing the abstract and full stage screenings to be conducted in parallel: in the abstract stage screening researchers identify intermediate relevant documents as early as possible from ranked documents, and they can start the full document screening on the intermediate documents, while the rest documents are still in the abstract screening. And also they are likely to be irrelevant. In this work, we focus on improving the abstract stage screening in screening prioritization.

Seed-driven document ranking (SDR) approach uses a known relevant document (seed) as a query in screening prioritization. Existing work in SDR seeks ways to measure different terms weights in a query document and combine them in a retrieval model to compute ranking scores~\cite{Seed_driven_SDR}. In this work, we study an alternative way in SDR: ranking by similarities to a query document. We propose a simple document matching measure named Mirror Matching (\mmatch, for short), which leverages a common writing pattern in abstracts in medical literature. Abstract text is comprised of background, method, result and conclusion in order. The proposed measure matches terms in two documents with respect to their positions, so that it enables to match terms that are used in the same section.

The proposed measure is motivated by observations on positional patterns of PICO elements. PICO elements indicate four elements, \textit{P}atient, \textit{I}ntervention, \textit{C}omparator, and \textit{O}utcome, and they are the main components of relevance conditions in systematic reviews. Hence,  the relevance of documents is mainly determined by PICO elements in documents. A  document must have the same PICO elements as a query document does in order to be a relevant document. For example, if a term `insulin' is in \textit{O} element in a document and it is in \textit{I} element in a query, the document can't be a relevant document, even though both contain the same term. In other words, terms must be compared with respect to each of PICO elements in order to identify relevant documents. We observed that each PICO element appears with different patterns in abstract texts. For example, P element frequently appears in the beginning of document, and most O element terms appear at the end of document.
Using positional information is a relaxed and cost-effective way to match terms in regard to PICO elements, compared to specifically identifying PICO element terms, which is not a simple task~\cite{Prioritization_PICO_Recognition,PICOspanAgreement}.

\mmatch is a document similarity measure, inspired by the asymmetric scoring in retrieval models (\ie BM25). We discuss the connections and differences between \mmatch and existing document similarity measures and retrieval models. We conduct experiments on CLEF 2019 eHealth Task 2 TAR dataset and compare \mmatch to them as well.  In recent years, neural retrieval models gain significant attention for matching texts~\cite{NeurlIR_survey_Maarten,NeuralIR_survey_Microsoft,NeuralIR_survey_umass}. We also evaluate \mmatch in comparison to several neural IR models. The evaluation  results show that the proposed model outperforms all baseline models on Average Precision and Precision-focused metrics.

The main contributions of this work are as follows. First, we study the SDR task in the context of the document matching for the first time and propose the effective document matching measure for medical abstracts. It allows SR researchers to find multiple relevant documents just after screening a couple of top-ranked documents. Second, the proposed approach is extensively evaluated compared to a wide range of baseline models from document similarity measures to classical to neural IR models. Lastly, we discuss the strength and the weakness of \mmatch and suggest ways to utilize \mmatch at most in practice while minimizing its limitation.

\section{Related Work}
\label{sec:Related_work}
\paratitle{Improving the screening process.} Several approaches have been proposed to reduce the costs of the expensive screening step in SRs~\cite{SR_survey}. They are mainly divided into two categories based on its end goals. One category is to directly help the screening step to be more effective, either by decreasing the number of candidate documents to be screened or by reducing the number of SR experts needed. This direct goal is pursued by training a classification model. However, adopting a trained classifier is prone to fail to find \textit{all} relevant documents, because it is hard to ensure the nearly perfect level of accuracy and also different SRs are meant to answer different clinical questions. That is, each SR has its own topic and it is hard to learn a general classification model for all SRs. Moreover, the small number of relevant documents in each SR makes training a model more challenging. With this regards, several studies suggest active learning setting through relevance feedback~\cite{CAL_waterloo,wallace_AL}.

The other category is a relatively indirect way to improve the screening process. While the number of documents to be screened remains unchanged, it assists SR experts during the manual screening process. An advantage of this approach is not to imperil finding \textit{all} relevant documents. Screening prioritization is an approach in this category. As aforementioned, it makes the screening process efficient by allowing abstract screening and full document screening to be carried out simultaneously. Another approach is to visualize groups of similar candidate documents and provide their topics. In this way, SR experts are able to screen similar documents together, instead of documents in a random order.
We refer to~\cite{SR_survey} for more details.

\paratitle{Ranking documents using a query document.}
Using an example document as a query is an important research problem in various domains, such as patent search and literature search~\cite{QueryByDoc:WSDM:2009,QueryByDoc:SIGIR11:Q_decomposition,QueryByDoc:ICTIR15:patent_search,QueryByDoc:SIGIR14:Q_suggestion,QueryByDoc_Lewis_Retrieval_Richness,Seed_driven_SDR}. Most current solutions first transform a long query to a short query, and then apply a retrieval function. As most retrieval functions are optimized for short queries, verbose queries often become a challenge in computation and query understanding~\cite{Handbook_verbose_queries}.

In screening prioritization, formulating a short and meaningful query from a known relevant document is non-trivial  because of the detailed relevance conditions. There are a few possible approaches: keyword extraction and PICO element extraction. Keyword extraction techniques can be adopted, but they are prone to lose detailed yet important relevance conditions such as age or gender of patients in P-element. Extracting medical concepts can be an alternative~\cite{Seed_driven_SDR}, but this approach has to rely on the accuracy of toolkits for medical concept extraction. PICO element extraction can be employed to extract relevance conditions appeared in a query document~\cite{PICOspanAgreement,RetreivalwPICO_ECIR2010,Prioritization_PICO_Recognition}. Extracting PICO elements without loss of relevance information therefore becomes a challenge.

\textit{Document matching} approaches have been less explored in ranking by a query document. Computing pairwise document matching scores in large-scale corpora online is costly. Several studies propose ways to improve the efficiency in pairwise similarity search problems~\cite{EfficiencyDocPairSim:AllPair_Similarity_Search,EfficiencyDocPairSim:jimmylinSIGIR09:bruteForce,EfficiencyDocPairSim:jimmylin:acl09short:mapreduce}.
However, in screening prioritization, candidate documents are typically in the scale of several thousands documents. In the  CLEF TAR 2019 dataset, the average number of candidate documents is about 6,000. The computation on a few thousand documents is small compared to the costs of manual screening which takes place on prioritized documents for weeks.  The effectiveness of prioritizing documents is more crucial than the efficiency in improving the manual screening process.

\section{Mirror Matching}
\label{sec:Mirror_Matching}

We focus on screening prioritization using a known relevant document in systematic reviews. The task is to rank candidate documents where the unknown relevant documents are aimed to be at top positions. We view this task in the context of document matching. Given a document pair (\ie the known relevant document and a candidate document), we compute a matching score between the two documents and rank candidate documents according to their matching scores with the known relevant document. Existing pairwise document similarity measures can also be used to compute the matching scores. For easy presentation, in the following, we refer the known relevant document as a query document.

\subsection{Overview}
\label{subsec:Proposed_Overview}

Mirror Matching is a simple two-way semantic matching for a pair of documents. It examines (i) whether a candidate document is the best reflection of a query document, and (ii) whether a query document is the best reflection of a candidate document. Specifically, given a query document $Q$ and a candidate document $D$, it computes two scores: $sc_{Q \shortrightarrow D}$ and $sc_{D \shortrightarrow Q}$
\begin{equation}
\label{Equation:MMSumOfTwo}
\text{M-Match}(Q,D)=sc_{Q \shortrightarrow D}+sc_{D \shortrightarrow Q}
\end{equation}
where $sc_{Q \shortrightarrow D}$ is a  candidate document score computed by matching terms in $Q$ to terms in $D$; $sc_{D \shortrightarrow Q}$ is a query document score computed by matching terms in $D$ to terms in $Q$.

As shown above, in \mmatch a matching score is computed in two directions (from $Q$ to $D$ and from $D$ to $Q$). Hence, we call it two-way matching. We design the two scores to affect the final matching score $\text{M-Match}(Q,D)$ equally.\footnote{We experiment with alternative combinations,  multiplication of two scores, and observe similar performance.}

Before explaining further details of \mmatch, we present an analysis on positions of PICO elements. Our analysis shows that  PICO elements appear with  positional patterns in medical literature writing. This analysis serves as the base to design our \mmatch to be position-based when computing $sc_{Q \shortrightarrow D}$ and $sc_{D \shortrightarrow Q}$.

\subsection{Analysis: Distributions of PICO Elements}
\label{subsec:PICO_Position}

Medical scientific abstracts are carefully written by authors to provide comprehensive overview of a full scientific literature. 
They summarize the major aspects of the entire literature in a prescribed sequence that includes background/purpose, method, results, and conclusion.
For example, American National Standard Institute advises authors to follow the rhetorical sequence when they write abstracts~\cite{AbstractStructure}. 
Furthermore, the ordered elements are explicitly indicated with headings in some abstracts.

Our hypothesis is that PICO words appear in patterns in medical scientific abstracts.
We study the positional distributions of PICO elements using EBM-NLP dataset~\cite{PICO_dataset}. EBM-NLP dataset provides manually annotated PICO spans and every token in documents is labeled. The interests on positions of PICO elements are not new~\cite{EMNLP_2010_PICOposition_retrievalmodel,PICO_position_Boudin_Nie_first_work}. Existing studies detect PICO elements using string matching with PICO keywords and report coarse-grained (\eg chunk-level or sentence-level) distributions of PICO elements. However, a string matching occurs common vocabulary mismatch between words describing PICO elements in medical literature and the PICO keywords.
Here, we cover various surface forms of PICO elements such as paraphrasing and synonyms using the manual PICO annotations available in EBM-NLP dataset. Besides, this analysis shows distributions of PICO elements in a fine-grained level (\eg words).

More specifically, EBM-NLP dataset contains about 5,000 PubMed documents (title and abstract only) with PICO span annotations. The dataset combines C-element and I-element as one label, I-element. Hence, every single-word term has one of the four labels: \underline{P}atient, \underline{I}ntervention, \underline{O}utcome, or \underline{N}one. When PIO elements are described in a long phrase, consecutive terms are labeled with the same labels. EBM-NLP consists of the labels by multiple annotators and the combined label that aggregates the labels by multiple    annotators. We use the combined version of labels.

\begin{figure}[t]
\includegraphics[width=0.75\linewidth]{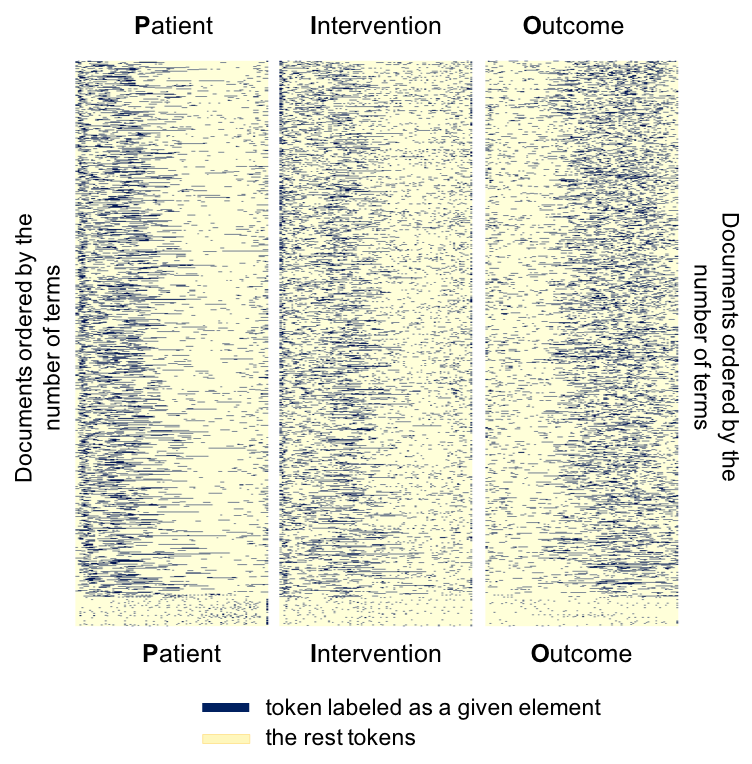}%
\caption{Distributions of Patient, Intervention, and Outcome elements in medical literature (title and abstract). Each row indicates a document. Within a row dark blocks are terms labeled with a given element (best viewed in color).}
\label{fig:PICO_Position}
\end{figure}

Figure~\ref{fig:PICO_Position} shows distributions of P-, I-, and O-elements. Documents are sorted by their lengths in number of words, in descending order. About 250 documents in EBM-NLP dataset have only a title, so that their lengths are much shorter than the rest and hard to show any patterns (see the bottom of the figure). In Figure~\ref{fig:PICO_Position} each row is a document, consisting of a sequence of terms.  Terms in different colors  indicate with or without a corresponding element label. Since documents have different lengths, the positions of terms within a row represent their relative positions in the document.

As shown in Figure~\ref{fig:PICO_Position}, P-element often appears at the beginning of documents where a title, \textit{background} and/or \textit{purpose} are likely to be stated. In the second half of documents it seldom appears.  Next, I-element tends to appear in the beginning of documents, similar to P, but its occurrence in the first half of documents  is not as apparent as that of P-element. Instead, I-element constantly appears in the second half of documents. Intervention is the main topic in medical articles, so that all parts (\textit{background/purpose}, \textit{method}, \textit{results}, and \textit{conclusion}) in documents are likely to involve I-element. Another distinct characteristic of I-element is that it is mostly short keywords, while P-element tends to be phrases. Observe that the difference in lengths of labeled words in P- and I-elements. Lastly, O-element  often appears at the end of documents  where \textit{results} and \textit{conclusion} are. More generally, O-element is mainly found in the second half of documents.

This analysis shows that each PICO element presents its own distributional pattern in documents and it is similar to the findings reported in~\cite{EMNLP_2010_PICOposition_retrievalmodel}. Hence, terms in same PICO elements are likely to be in similar parts (\eg beginning or end) of documents.  We  therefore design \mmatch to be position-aware by matching terms at similar positions. In that way, we attempt to match terms that are likely to be used for same PICO elements.  In the next section, we explain details of \mmatch.

\subsection{Mirror Matching: Two-Way Matching}
\label{subsec:Two_Way_Matching}
\begin{figure}[t] \includegraphics[width=0.75\linewidth]{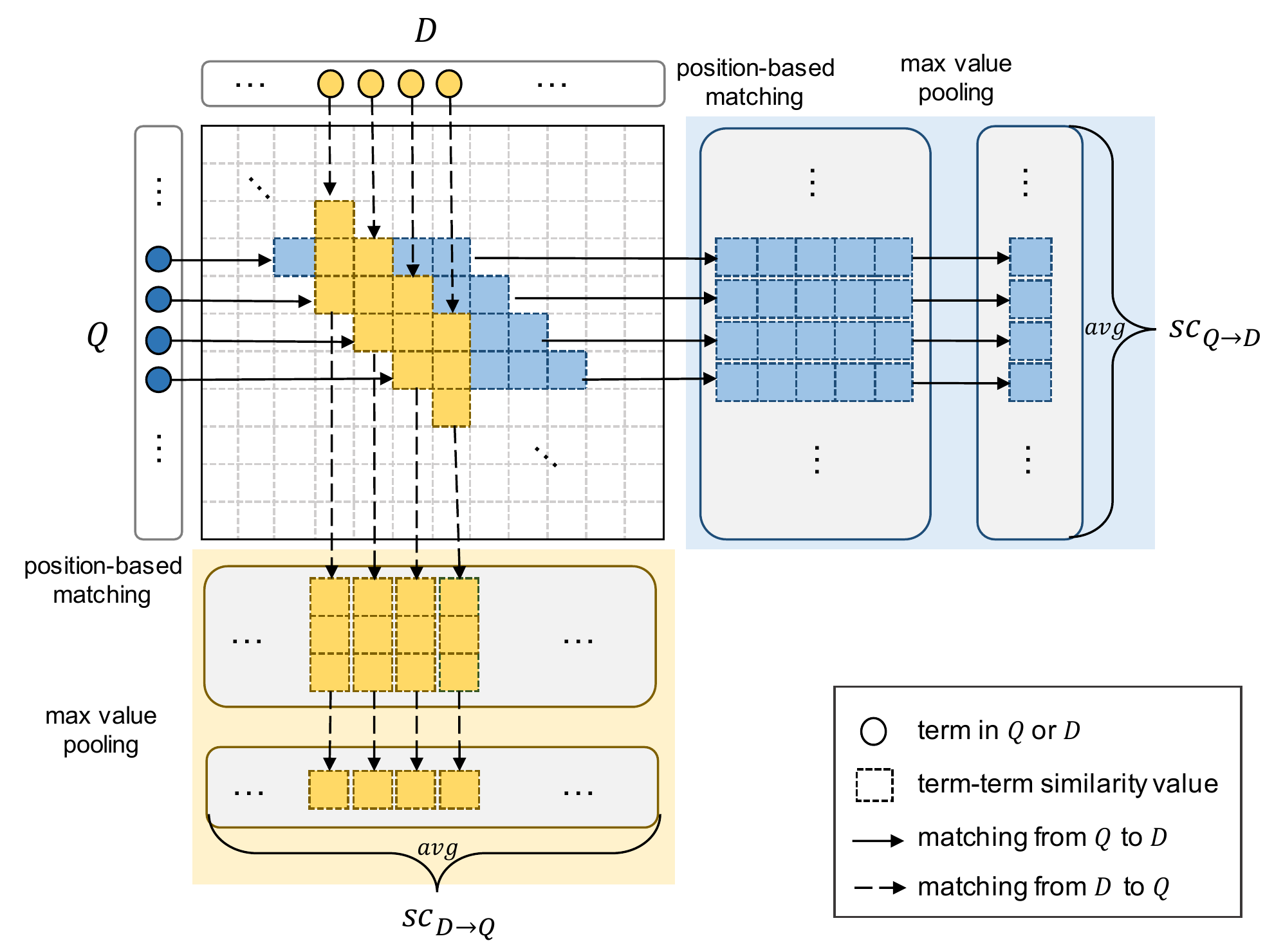}%
\caption{Mirror Matching uses two one-way interactions: from a query document $Q$ to a candidate document $D$ and from a candidate document $D$ to a query document $Q$. Term-term matchings are based on positional information and among them a maximum matching value is taken.}
\label{fig:MM_overview}
\end{figure}

Figure~\ref{fig:MM_overview} visualizes Mirror Matching, consisting of two one-way matchings. Each one-way matching is position-based matching between terms in a query document and terms in a candidate document.

We detail Mirror Matching in Algorithm~\ref{alg:Mirror_Matching}. Let $Q$ and $D$ denote a query document and a candidate document, respectively. A query document $Q$ consists of a sequence of query terms, $\langle q_1, q_2, \cdots, q_{|Q|} \rangle$, and a candidate document $D$ consists of a sequence of document terms $\langle t_1, t_2, \cdots, t_{|D|} \rangle$. The subscription of each term indicates its position in a given document.  In Mirror Matching, a matching score between $Q$ and $D$, $\text{M-Match}(Q,D)$, is the sum of two one-way scores: $sc_{Q \shortrightarrow D}$ and $sc_{D \shortrightarrow Q}$, as presented in Equation~\ref{Equation:MMSumOfTwo}. The concept behind the two one-way scores is identical. Hence, we only explain the computation of $sc_{Q \shortrightarrow D}$ as a representative.
\begin{algorithm}
\caption{Mirror Matching Overview}
\label{alg:Mirror_Matching}
\begin{algorithmic}[1]
\Require{Query document $Q=\langle q_1,q_2, \cdots, q_{\left|Q\right|}\rangle$,\newline Candidate document $D=\langle t_1,t_2, \cdots, t_{\left|D\right|}\rangle$}
\Ensure{M-Match$\left(Q,D\right)$ score between $Q$ and $D$,}
\Statex
\State{initialize $sc_{Q \shortrightarrow D}$ and $sc_{D \shortrightarrow Q}$ to zero}
\For{$i \gets 1$ to $\left|Q\right|$}\Comment{start one-way matching from $Q$ to $D$}
    \State{initialize $S_{temp}$ an empty list/array}
    \For{each $j \in$ \texttt{matching\_position$\left(\lambda, i,\left|Q\right|,\left|D\right|\right)$}}                         \State {$S_{temp}.append\left(sim\left(q_i,t_j\right)\right)$}
    \EndFor
    \State{$sc_{Q \shortrightarrow D} = sc_{Q \shortrightarrow D} + max\left(S_{temp}\right)$}
\EndFor
\State{$sc_{Q \shortrightarrow D}=sc_{Q \shortrightarrow D}/\left|Q\right|$}
\For{$j \gets 1$ to $\left|D\right|$}\Comment{start one-way matching from $D$ to $Q$}
    \State{initialize $S_{temp}$ an empty list/array}
    \For{each $i \in$ \texttt{matching\_position$\left(\lambda, j,\left|D\right|,\left|Q\right|\right)$}}                         \State {$S_{temp}.append\left(sim\left(q_i,t_j\right)\right)$}
    \EndFor
    \State{$sc_{D \shortrightarrow Q}=sc_{D \shortrightarrow Q}+max\left(S_{temp}\right)$}
\EndFor
\State{$sc_{D \shortrightarrow Q}=sc_{D \shortrightarrow Q}/\left|D\right|$}
\State{$M-Match(Q,D) = sc_{Q \shortrightarrow D}+sc_{D \shortrightarrow Q}$}\Comment{final matching score}
\State \Return {$M-Match(Q,D)$}
\end{algorithmic}
\end{algorithm}

The score $sc_{Q \shortrightarrow D}$ is the average of scores computed on each query term. For each term $q_i$, a score is computed using its positional information by finding a term in $D$ which has a similar position in $D$ as the position of $q_i$ in $Q$ and is also the best semantic matching to $q_i$. Therefore, we first find terms in $D$ that are in similar relative positions to given $q_i$, and then identify the best matching term among them.

More specifically, we find terms in $D$ that are in similar positions to $q_i$ using a window size. The position of $q_i$ is transformed into a relative position in $D$, which is $|D|\cdot i/|Q|$. Note that $i/|Q| \in (0, 1]$.  We then set a range of positions using a window size. As document lengths vary, we define a window size proportional to the length of a given document using a hyperparameter $\lambda$. For example, if  $\lambda=0.3$, then $30\%$ of the document length is set as a window size. Hence, the range for similar positions in $D$ given $q_i$ is calculated as $(|D|\cdot i/|Q|)\pm \lambda\cdot|D|$. Different signs denote preceding and following directions from a relative position. The range includes the beginning and ending numbers of the interval. In Algorithm~\ref{alg:Mirror_Matching}, \texttt{matching\_position} denotes the procedure of finding matching terms that are in similar positions. Next, each candidate term in the positional range is matched to a query term $q_i$. A matching between two terms is a cosine similarity of their word embeddings. Among the cosine similarity values, a maximum value is chosen ($1$-max pooling) to exploit the most informative matching signal. Word embeddings allow semantic matching, not limited to surface forms of terms.

In summary, given $q_i\in Q$ we find the best matching term $t_j\in D$ within the matching position in $D$ and the term-term matching score between $q_i$ and $t_j$ is computed by cosine similarity in word embedding space. In this sense, our model is able to consider both PICO elements matching and semantic matching. Finally $sc_{Q \shortrightarrow D}$ is the average of scores obtained from each query term $q_i\in Q$.

The computational cost of $sc_{Q \shortrightarrow D}$ is governed by the number of terms in $Q$, the number of terms in  $D$, and $\lambda$ that determines a window size. The estimation of $sc_{Q \shortrightarrow D}$ has the time complexity  of $O(nk)$ where $n$ is the number of terms in $Q$ (\ie $|Q|$) and $k$ is the number of terms within matching positions of $D$.

Likewise, $sc_{D \shortrightarrow Q}$ is the average of scores computed on each term in $D$. It is worth noting that the two one-way scores, $sc_{Q \shortrightarrow D}$ and $sc_{D \shortrightarrow Q}$, are \textbf{\textit{asymmetric}} as they are computed with respect to the terms in query and candidate documents, respectively. In our evaluation, we set $\lambda$ to $0.35$.

\subsection{Discussion}
\label{subsec:Model_Discussion}

We discuss the connections and differences of \mmatch with ad hoc retrieval models and document similarity measures.

\paratitle{Relationship with ad hoc retrieval models.} In ad hoc retrieval, typically a query is short and a document is long. Since a given query is a primary information source for relevance, retrieval models (\eg BM25) estimate document scores computed on each term in query (one-way scoring)~\cite{chengxiang_zhai_handbook}. In our problem setting, a query is a (known relevant) document, same as a candidate document. Besides, if a candidate document happens to be a relevant document, it can be a query document as well. Hence, our \mmatch model employs a two-way scoring which demonstrates both of them can be a query document. The connection to retrieval models is that our model maintains the asymmetry in the one-way scorings, similar to that in retrieval models. We discuss the purpose in comparison with document similarity measures.

\paratitle{Relationship with document similarity measures.} Common document similarity measures (\eg cosine similarity between two document representations) are symmetric. Given two documents $d_1$ and $d_2$, $similarity(d_1, d_2)$ is the same as $similarity(d_2, d_1)$. Unlike symmetric document similarity measures, our proposed \mmatch employs two asymmetric scores. We design asymmetry in the scores in order to find the most similar documents, from both query and candidate documents' perspective. We use an example to further illustrate this point.

Let us assume that $d_1$ and $d_2$ are similar, but $d_2$ includes more detailed content while fully covering all content of $d_1$. Given the two documents, existing document similarity measures are likely to estimate a high similarity value since they are similar as a whole. In the context of document matching, when finding documents matching to $d_1$, $d_2$ can be a good document for $d_1$ because all content of $d_1$ is fully covered in $d_2$. However, when finding documents matching to $d_2$, $d_1$ may not be a good document for $d_2$, as $d_1$ lacks the detailed content in $d_2$. The asymmetric scores in \mmatch give a lower score in this case, especially in the matching score from $d_2$ to $d_1$.

\section{Evaluation Settings}
\label{sec:Eval_set}

\subsection{SR Dataset}
\label{subsec:Data}

We conduct our evaluation on CLEF2019 dataset. This dataset contains 79 diagnostic test accuracy (DTA) SRs, 20 Intervention, 2 Qualitative, and 1 Prognosis SRs  in CLEF eHealth 2019 Technologically Assisted Reviews (TAR)~\cite{clef_19_overview}. CLEF2019 is currently the most comprehensive dataset for screening prioritization.  We use 79 DTA SRs in the evaluation. DTA SRs are considered as a difficult type of SRs to automatically identify relevant documents because of their broad range of topics. The other types of SRs in CLEF2019 are much fewer than DTA SRs, and are not included\footnote{Note that for DTA SRs, the original CLEF2019 consists of 80 SRs, which are 72 SRs from CLEF2018 and 8 new SRs. One new SR (id: CD011686) is already in CLEF2018 and it has different relevance labels in CLEF2018 and in CLEF2019. We use the labels from CLEF2018.}. Training/testing data splits are defined in CLEF2019 but we use all SRs since our model does not need training.

CLEF2019 provides document identifiers (PMIDs) of candidate documents with their two relevance labels (from abstract screening and full document screening). We use the label from full document screening in the evaluation because these labels indicate the final relevant and non-relevant documents in SRs. From a public MEDLINE 2019 corpus, we obtain document contents (title, abstract, and MeSH keywords) using document identifiers. For the detailed statistics of the dataset, we refer to~\cite{clef_19_overview}.

Any relevant document can be the known document in a SR. We use every relevant document as a query and rank the rest candidate documents. In CLEF2019, a few SRs have relevant documents fewer than 2. In this setting, as one relevant document is used as a query, they are not included in the evaluation.

\paratitle{Pre-processing.}
In medical literature, short forms (\eg abbreviations or acronyms) of long medical terminologies are often used for effective communication. 
However, there is no standard rule for constructing short forms. 
Hence, a long medical term is written in different short forms across different documents. 
To increase consistency within and between documents, we identify short forms and their long forms using a simple regular expression, and transform the short forms into the long form. We also detect three kinds of numeric expressions and normalize them into INT, FLOAT, and PERCENT. Lastly, stop words are removed.

\paratitle{Word Embeddings.}
We use word embeddings locally trained on candidate documents per SRs. 
As candidate documents are retrieved by Boolean keyword queries derived from  relevance conditions, they are a collection of similar documents. 
Multiple studies report that large corpora do not necessarily produce good word embeddings in  biomedical domain, compared to a smaller but topic-specific collection~\cite{CliEmbed_eval_Howto_train_embeddings_biomedical,CliEmbed_eval_LargeCorpus_not_enhance}.
A similar finding has also been reported in a general domain problem~\cite{Local_embed_Mitra}. 
Our empirical results also show that local word embeddings are more effective than global word embeddings trained on a corpus (we report details in Section~\ref{subsubsec:Analysis_Embeddings}). 
We use skip-gram with negative sampling implemented in gensim. The dimension of word embeddings and a window size are set to 300 and 7, respectively and the minimum word occurrence is set to 5. The default settings are used unless it is specified.

\subsection{Baselines}
\label{subsec:Baselines}
The evaluation is done with ad hoc retrieval models and document similarity measures. Candidate documents are prioritized by retrieval scores or similarity scores.

\paratitle {Retrieval models.} Two classical retrieval models (BM25 and query likelihood model) and a retrieval model specifically designed for screening prioritization (SDR-BOC) are evaluated. \textbf{BM25} is a simple and strong retrieval model and it has continuously proved its effectiveness~\cite{Lin_neuralHype_SIGIR_short}. We use BM25 implementation in gensim. Query likelihood model (\textbf{QL}) is a language model approach and we experiment QL with JM smoothing and a smoothing parameter is empirically set to 0.2. \textbf{SDR-BOC} is a retrieval model proposed for screening prioritization~\cite{Seed_driven_SDR}. It is QL combined with an additional term weighting function. As suggested in the original paper,  we represent documents in bag-of-concepts (BOC) after extracting medical concepts.

\paratitle{Document similarity measures.} 
We examine document similarities using lexical and semantic matching signals.
The first measure is a cosine similarity with different document representations: TF-IDF, AvgEm-W, AvgEm-C, and ParaVec. \textbf{TF-IDF} is a bag-of-words document representation using the classic \textit{tf-idf} weighting scheme. \textbf{AvgEm-W} is the average of word embeddings of all words contained in a document. Likewise, \textbf{AvgEm-C} is the average of concept embeddings of all concepts contained in a document. A concept embedding is the average of word embeddings of all words contained in a medical concept. Paragraph vector (\textbf{ParaVec}) is a distributional document representation proposed by~\cite{PV_DBOW_para_vec}. We use PV-DBOW in gensim.
In addition, we tested different similarity functions such as Euclidean distance in pilot experiments, but cosine similarity showed the highest performance.

We also evaluate document similarity measures proposed by existing studies.
\textbf{TF-inner}~\cite{EfficiencyDocPairSim:jimmylinSIGIR09:bruteForce} computes an inner product of two bag-of-words document representations using \textit{tf} weighting.
\textbf{OK}~\cite{EfficiencyDocPairSim:cikm13:OK} is motivated by BM25 and computes a document similarity using BM25 weights. Word mover's distance (\textbf{WMD})~\cite{WMD_wordmoverdist}  estimates a distance between two documents defined as a cost when all words in one document move to words in the other document. WMD is a symmetric distance metric and its negative value is considered as a similarity. 
\cite{CIKM_wmd_IR} showed that WMD can be applied to ad hoc retrieval.

AvgEm-W, AvgEm-C, and WMD use the same word embeddings as \mmatch. In SDR-BOC and AvgEm-C,  QuickUMLS~\cite{QuickUMLS} is used to extract medical concepts.

\subsection{Evaluation Metrics}
\label{subsec:Eval_Measures}
We use evaluation metrics commonly adopted in screening prioritization task: average precision, precision at $k$, recall at $k$, and work saved over sampling at recall 100\%.
Average precision (AP) is a metric combining a recall component to precision-focused metric by averaging all values of precision at each relevant document. Precision at $k$ (Pr@$k$) is the ratio of relevant documents out of top $k$ documents, and recall at $k$ (Re@$k$) is the ratio of relevant documents in top $k$ documents out of the total number of relevant documents $k$ is set to 10, 20, and 30.  Work saved over sampling at recall 100\% (WSS100) was introduced by \cite{cohenWSS} and it measures the ratio of documents that are ranked lower than a last relevant document out of the total candidate documents. WSS100 estimates the workload to be reduced in the manual screening with the assumption that after a last relevant document is found the rest unlabeled documents are not screened. 
Finally, each SR is evaluated by these evaluation metrics and then the mean over all SRs is reported.

\section{Results and Discussion}
\label{sec:Results}

In this section, we aim to answer three research questions.

\paratitle{RQ1.} Is \mmatch more effective than retrieval models and/or document similarity measures for screening prioritization using a query document?

\paratitle{RQ2.} To what extent does each component (\eg two-way matching, position-aware matching) in \mmatch contribute to its performance?

\paratitle{RQ3.} Regarding recent success of neural IR models in matching two texts, how do neural models perform in this task?

\begin{table*}
\centering
\caption{Comparison between Mirror Matching and the baselines. The best model is in bold and the second best is underlined in each metric. $^*$ denotes statistically significant performance of the best model compared to a given model (p-value $<0.01$).}
\label{tab:Overall_Result}
\begin{tabular}{lllllllll}
\midrule
\textbf{Models} & \textbf{AP}   & \textbf{Pr@10}     & \textbf{Pr@20}     & \textbf{Pr@30}     & \textbf{Re@10}   & \textbf{Re@20}    & \textbf{Re@30}   & \textbf{WSS100}\\
\midrule
BM25 & 0.169     & 0.209     & 0.173     & 0.151     & \textbf{0.178}     & \underline{0.246}     & \underline{0.299}    & 0.437$^*$\\
QL  & 0.155     & 0.189     & 0.158     & 0.139     & 0.170      & 0.235     & 0.279    & 0.466$^*$  \\
SDR-BOC & 0.172     & 0.209     & 0.174     & 0.152     & 0.176     & \textbf{0.251}     & \textbf{0.303}    & 0.541$^*$\\
\midrule
TF-IDF  & 0.180     & 0.208     & 0.173     & 0.153     & 0.127     & 0.192     & 0.243    & \underline{0.690} \\
AvgEm-W  & 0.176     & 0.201     & 0.167     & 0.147     & \underline{0.177}     & 0.242     & 0.290     & 0.530$^*$\\
AvgEm-C  & 0.167     & 0.190      & 0.162     & 0.145     & 0.110$^*$      & 0.175     & 0.228    & \textbf{0.692 } \\
ParaVec  & 0.139$^*$     & 0.164$^*$     & 0.141$^*$     & 0.126$^*$     & 0.093$^*$     & 0.153$^*$     & 0.198$^*$    & 0.598  \\
\midrule
TF-inner & 0.119$^*$	&0.136$^*$	&0.115$^*$	&0.102$^*$	&0.112	&0.172	&0.214	&0.477$^*$	\\
OK      & 0.113$^*$	&0.113$^*$	&0.107$^*$	&0.100$^*$	&0.066$^*$	&0.125$^*$	&0.173$^*$	&0.650	 \\
WMD & \underline{0.203}     & \underline{0.232}     & \underline{0.194}     & \underline{0.170}      & 0.141     & 0.225     & 0.272    & 0.592 \\
\midrule
\mmatch & \textbf{0.220}      & \textbf{0.247}     & \textbf{0.207}     & \textbf{0.183}     & 0.149     & 0.234     & 0.293    & 0.687 \\
\midrule
\textit{Rank ($\Delta$)}    & 1 (+0.017) & 1 (+0.015) & 1 (+0.013) & 1 (+0.013) & 5 (-0.029) & 5 (-0.017) & 3 (-0.010) & 3 (-0.005) \\
\midrule
\end{tabular}
\end{table*}

\subsection{Overall Results}
\label{subsec:Overall_Res}

First, to answer \textbf{RQ1} we present the overall results of the proposed model and the baselines in Table ~\ref{tab:Overall_Result}. We also report the rank of the proposed model among all models and a performance gap between the proposed model and the best baseline model in the last row. A positive/negative sign in the performance gap indicates win/loss of the proposed model.

\mmatch outperforms all baseline models in terms of AP and Pr@$k$.  
Two-way matching scores effectively rank relevant documents higher than non-relevant documents.
WMD is the second best model in AP and Pr@$k$. While both \mmatch and WMD use semantic term matching, a key difference is that \mmatch  matches terms using positional information to match terms with same PICO elements. 
On the other hand, WMD matches terms by minimizing the cases of matching same terms multiple times regardless positions and it also leads to higher computation cost than \mmatch.

Two retrieval models, BM25 and SDR-BOC, show the highest Re@$k$ values and AvgEm-C has the highest WSS100 value. Indicated in the last row of Table~\ref{tab:Overall_Result},  \mmatch follows behind these models with a small gap.
The high performance of SDR-BOC and AvgEm-C supports that bag-of-medical concepts (BOC) is an effective representation for medical documents~\cite{Koopman:UsingConcepts,Seed_driven_SDR,PLOS_18_concept_extraction_vs_DL}. Bag-of-words (BOW) and bag-of-concepts (BOC) represent documents with their own advantages and disadvantages. BOW takes full advantage of all words in a document, despite sparsity.   On the other hand, BOC relies on essential medical terminologies with a risk of losing contextual information in non-medical words.

We claim that the strong AP and Pr@$k$ of \mmatch are particularly useful in the beginning of manual screening process.
The higher AP and Pr@$k$ values represent the more relevant documents to be found when SR researchers start screening a handful of top-ranked (\eg top 10) documents.
When more relevant documents are identified early, they can further improve the rankings of remaining unlabeled documents.

\subsubsection{Ablation Study}
\label{subsubsec:Ablation}

\begin{table}
\centering
\caption{AP of degraded models without key components of Mirror Matching.}
\label{tab:ablation_study}
\begin{tabular}{llc}
\midrule
\textbf{Model} & \textbf{AP} & \textbf{loss} \\
\midrule
\midrule
\mmatch w/o position     & $0.216$ & $-0.004$\\
\mmatch w/o two-way    & $0.204$& $-0.016$\\
\mmatch w/o \{two-way, position\}     &$0.196$  & $-0.024$\\
\midrule
\mmatch   &  $0.220$ & -\\
\midrule
\end{tabular}
\end{table}

We investigate the effect of major components in the proposed model: two-way matching mechanism and position-aware matching. Three degraded models are examined in an ablation test.

\textbf{\mmatch w/o position} removes positional information in finding the maximum term matching similarity. In other words, a max-pooling is applied to the entire row/column in Figure~\ref{fig:MM_overview}. \textbf{\mmatch w/o two-way} removes two-way matching mechanism. A final matching score is $sc_{Q\shortrightarrow D}$, without $sc_{D\shortrightarrow Q}$.
\textbf{\mmatch w/o \{two-way, position\}} removes both components in Mirror Matching. A matching score is $sc_{Q\shortrightarrow D}$ and also it does not use positional information in max pooling.

Table~\ref{tab:ablation_study} presents performances of the degraded models compared to \mmatch. We observe that: (1) \textit{\mmatch w/o two-way} performs worse than \textit{\mmatch w/o position}, showing that incorporating two one-way scores has the bigger impact than position-aware matching.
(2) Performance of \textit{\mmatch w/o \{two-way, position\}} has a significant drop from \mmatch.
AP of \textit{\mmatch w/o \{two-way, position\}} is even lower than that of WMD reported in Table~\ref{tab:Overall_Result}. These two observations suggest that both components contribute to the proposed model. In the next section, we further study the two-way matching mechanism and investigate its major contribution to \mmatch.

\subsubsection{Analysis on Two-Way Matching}
\label{susubbsec:Analysis_Two_Way}
The key claim of two-way matching is to employ a score from candidate document's perspective ($sc_{D\shortrightarrow Q}$) as well as the commonly used score from query's perspective ($sc_{Q\shortrightarrow D}$). 
We study how $sc_{D\shortrightarrow Q}$ make a change in matching scores of relevant documents and eventually improves the rankings.

First we find  relevant and non-relevant document pairs that $sc_{Q\shortrightarrow D}$ of a non-relevant document has the smallest difference to $sc_{Q\shortrightarrow D}$ of a relevant document.
That is, the two documents in pairs are ranked next to each other without two-way matching (\ie $sc_{Q\shortrightarrow D}$ alone is a matching score). We then compare their $sc_{D\shortrightarrow Q}$. 
We calculate the portion of pairs such that $sc_{D\shortrightarrow Q}$ of the relevant document is greater than that of the non-relevant document within ranked candidate documents by a query document. We then report the average portion over all queries from all SRs in CLEF2019.\footnote{We also calculate the average portion via two levels: average over a SR and then average over all SRs. The resulting portion is consistent.} As a result, 63.5\% of the pairs have greater $sc_{D\shortrightarrow Q}$ of the relevant document than $sc_{D\shortrightarrow Q}$ of the non-relevant document. The result suggests that scores from candidate document's perspective $sc_{D\shortrightarrow Q}$ tend to promote relevant documents.

We note that the above analysis does not confirm the rise of relevant documents in a ranked document list as the rest of candidate documents are also involved in the total ranking. Nevertheless, this close analysis helps better understand the two-way matching in finding relevant documents.

\subsubsection{Impact of Window Size in Position-aware Matching}
\label{subsec:Analysis_Position}

\begin{figure}[t] \includegraphics[width=0.65\linewidth]{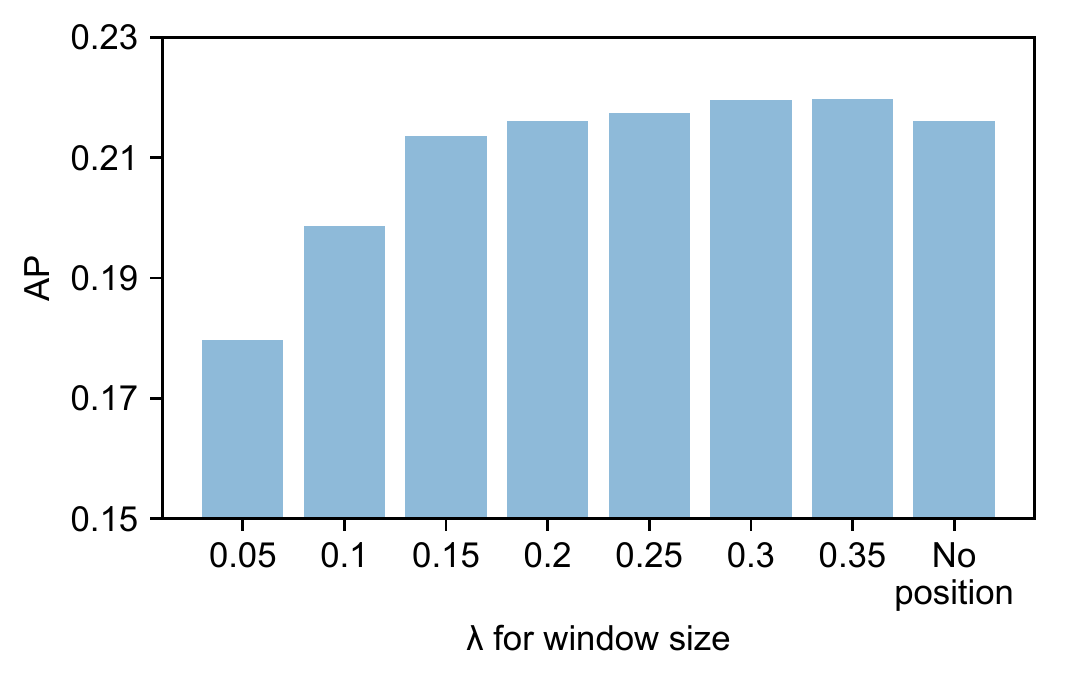}%
\caption{Impact of different ranges of matching positions. A range is twice of a window size, which is determined by $\lambda$.}
\label{fig:position_win_size}
\end{figure}

Position-aware matching in two one-way matchings helps match terms that are used as the same PICO element. Positional information of terms is incorporated with a window size determined by $\lambda$. Here we study the effect of different window sizes. We experiment different values of  $\lambda$ from $0.05$ to $0.35$ with a $0.05$ increment. We plot AP of \mmatch in Figure~\ref{fig:position_win_size}. 
Recall that a window size indicates one side range (either preceding or following) from a given term. When $\lambda=0.35$, for a term in the middle of document, it has a matching area of 70\% of a document length. 
Additionally, we display AP with no positional information \ie \textit{\mmatch w/o position}.

In Figure~\ref{fig:position_win_size}, AP gradually increases along the increase of window size. The highest performance is observed when $\lambda$ is 0.35. It is worth noting that when the model pinpoints terms with a very small window ($\lambda=0.05$), the performance deteriorates significantly. We believe the large drop originates from the different granularity of the rhetorical structure in documents. For example, some abstracts have much longer \textit{results} part than others.
Lastly, while using positional information of terms improves performance, the performance gain is relatively small compared to the case without using it. Referring to Figure~\ref{fig:PICO_Position}, the positional patterns of PICO elements can be seen in a collective manner among several thousands documents and the pattern in a pair of two documents may not be strong enough. Besides, using positional information is a soft way of incorporating PICO elements in matching as the PICO spans are not explicitly pre-detected~\cite{Prioritization_PICO_Recognition}.

\subsubsection{Impact of Word Embeddings}
\label{subsubsec:Analysis_Embeddings}
\begin{figure}[t] \includegraphics[width=0.65\linewidth]{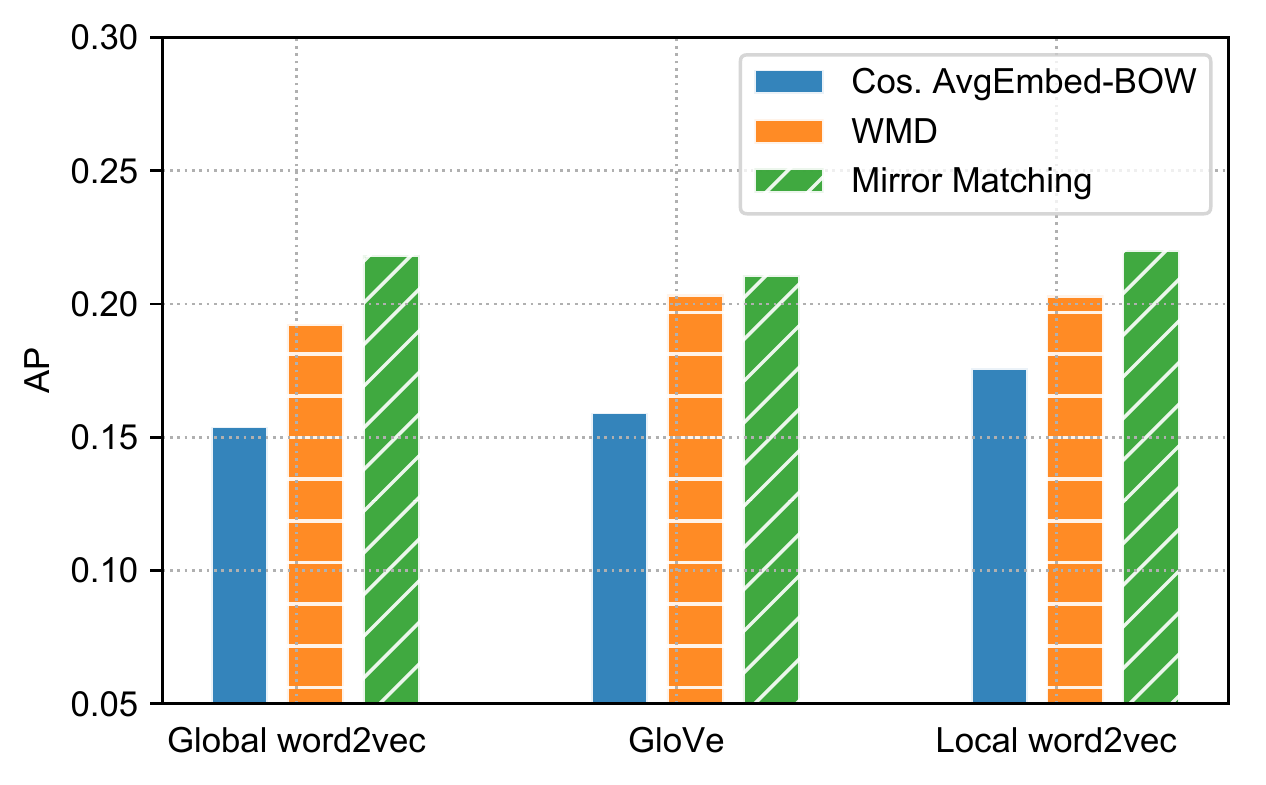}%
\caption{Impact of different word embeddings on cosine similarity with AvgEm-W, WMD, and \mmatch. (REDRAW A FIGURE WITH CHANGED MODEL NAMES (AvgEm-W)}
\label{fig:Effect_of_word_embed}
\end{figure}

The quality of word embeddings affects the performance of Mirror Matching, WMD, as well as cosine similarity over AvgEmb-W and AvgEmb-C. We now examine the impact of different word embeddings on AvgEmb-W, WMD, and the proposed model. We evaluate two  word embeddings pre-trained on large corpora: medical domain word2vec\footnote{\label{fn:Global_w2v} http://bio.nlplab.org/} (Global word2vec) and  GloVe\footnote{https://nlp.stanford.edu/projects/glove/} and also present the earlier evaluation with local word embeddings trained on candidate documents (Local word2vec). As shown in Figure~\ref{fig:Effect_of_word_embed}, in comparison of different word embeddings, locally trained word embeddings consistently  show the best performance in all three models compared to the other two embeddings. Furthermore, \mmatch consistently outperforms the two baselines in all different kinds of word embeddings.

We also experiment contextualized word embeddings using pre-trained BioBERT (version 1.0 PubMed PMC)~\cite{BioBERT}. However, while \mmatch outperforms the other models, the overall performance in all three models is much lower than distributional word representations (\eg word2vec, GloVe). We presume that this is related to the vocabulary coverage of BioBERT with respect to our dataset. Moreover, to the best of our knowledge, it is unknown how to properly adopt BERT in a word-level intrinsic evaluation setting, instead of extrinsic evaluation settings (\ie downstream tasks). We leave this as our future work.

\subsection{Ranking Updates with Multiple Documents}
\label{subsec:updating_rankings}
Next, we simulate the real-world scenario when the proposed model is used in the manual screening process.
We assume the manual screening is conducted on prioritized documents (\ie evaluated in Table~\ref{tab:Overall_Result}).
When top 20 documents are screened, the rankings of the remaining unlabeled documents are updated with new relevant documents found in the top 20 documents.
We repeat this three times.

As the proposed model computes document-to-document matching scores, each of newly found relevant documents produces a new ranked list. 
Multiple ranked lists also have different ranges of ranking scores, so we combine them using rankings.
For example, for an unlabeled document, we get its rankings in multiple ranked lists and then the averaged ranking is set to its updated ranking. 
Unlabeled documents are ordered by updated rankings in ascending order.
The rank-wise combination allows multiple ranked lists carry equal weights in the updated ranked list.

We evaluate an updated list by AP and WSS100 and they evaluate different aspects of the list (details in Section~\ref{subsec:Eval_Measures}.
We compare the proposed model against TF-IDF and AvgEm-C which showed the higher WSS100 values than the proposed model in Table~\ref{tab:Overall_Result}.

\begin{table}
\centering
\caption{Simulating the manual screening process. A ranked list of unlabeled documents is updated when every top 20 documents are labeled, which is defined as one round.}
\label{tab:Result_Updating_Ranking}
\begin{tabular}{c|ccc|ccc}
\midrule
&\multicolumn{3}{c|}{\textbf{AP}} & \multicolumn{3}{c}{\textbf{WSS100}}\\
\midrule
round & 1 & 2 & 3 & 1 & 2 & 3 \\
\midrule
TF-IDF   & 0.213&	0.241&	0.252&	0.745&	0.767&	0.777 \\
AvgEm-C  & 0.178&	0.204&	0.239&	0.704&	0.721&	0.731  \\
M-Match & 0.259&	0.262&	0.279&	0.719&	0.728&	0.732   \\
\midrule
\end{tabular}
\end{table}

Table~\ref{tab:Result_Updating_Ranking} presents the evaluation results of the simulation in three rounds.
As more relevant documents are used as a query document, all three models continue to improve the rankings for the unlabeled documents. \mmatch shows the highest AP in all three rounds and it is consistent with the high AP in the initial ranking results shown in Table~\ref{tab:Overall_Result}.  
On the other hand, the proposed model has the second best WSS100, following TF-IDF. The proposed model is not the most effective one to find a last relevant document.

\paratitle{Limitations.}
We believe the weak WSS100 in \mmatch is closely related to \textit{topic specificity/broadness of SRs}.
When a SR has broad topic, its relevant documents may share relatively lower similarities (\eg relevant documents cover different parts of broad relevance conditions).
In this case, it is hard for the proposed model to effectively find all relevant documents (\ie including a last relevant document).  
To examine the relation with topic specificity of SRs, we conduct a small analysis.

We assume that when the proposed model monotonically decreases WSS values over the three rounds for a SR, the proposed model does not work well for that SR to find all relevant documents. 
To estimate topic specificity, we use the minimum value in pairwise similarities between relevant documents in a SR. 
If a SR has specific topic, all relevant document pairs can have a high similarity and the minimum value is high as well.
Hence, the minimum value can reflect topic specificity of SR.

The analysis shows that the estimated topic specificity for SRs with increasing WSS values is 0.498. In contrast, the estimated topic specificity for SRs with decreasing WSS values is only 0.143. This result describes that SRs in which the proposed model is less effective tend to  have broad topic.

\paratitle{Suggestions.}
We have discussed the strength and the weakness of \mmatch: it is effective to find relevant documents at the top, but less effective to find a last relevant document (\ie a document less similar to a query document).
Therefore, adopting \mmatch in the beginning of the manual screening is particularly useful. 
It allows to identify more relevant documents after the small number of documents are screened.
Then, we suggest to combine/switch \mmatch to another ranking model (\eg TF-IDF) for the better ranking of a last relevant document. 
A supervised approach can also be used since several relevant and non-relevant documents are available.

\subsection{Neural Models in Screening Prioritization}
\label{subsec:Neural_IR_Models}
We have extensively evaluated \mmatch with \textbf{RQ1} and \textbf{RQ2}. 
To provide a complete study, we test a few neural IR models and answer the last research question \textbf{RQ3}.
Similar to \mmatch, neural models also find a pair of texts that match each other, but we first need to train them.

\subsubsection{Train and Test Settings.}
Neural IR models, a supervised approach, typically require large amounts of training data.
It is difficult to train neural models specialized for individual SRs because of too few relevant documents in a SR.
We use train and test splits defined in CLEF2019, consisting of 72 and 7 SRs, respectively\footnote{A development dataset is not defined in CLEF2019}. 
For a development (dev) dataset, we randomly sample 7 SRs in the training 72 SRs.  
Because of the random sampling, we examine two sets of train and dev datasets with different SRs. Both evaluation results are reported.

We generate train/dev instances (document pairs) using relevant and under-sampled non-relevant documents within each SR: \{relevant - relevant\} pairs for positive instances and \{relevant - non-relevant\} for negative  instances. In total, the train and dev datasets contain 145,104 instances. In the test dataset, a query document is paired with every candidate document.

We use MatchZoo (version 2.1.0) toolkit~\cite{MatchZoo} in this experiment.  MatchZoo toolkit provides a collection of representative neural IR models and it has been gaining attention from research communities as a solid experimental platform. We evaluate two simple neural baseline models (Dense, BiLSTM) and another two neural IR models (Duet, MV-LSTM)  which share similar matching (learning) mechanisms with Mirror Matching.
\begin{itemize}
    \item \textbf{Dense} is a naive baseline. Word tokens are represented as one-hot vectors and they are concatenated and encoded using two layers of feedforward neural networks (FNNs).
    \item \textbf{BiLSTM} is bidirectional LSTMs (BiLSTM) in a siamese architecture.  After embedding lookup and encoding by BiLSTM, the last hidden states from documents in a pair are concatenated. Then, two layers of FNNs are added. 
    \item \textbf{Duet}~\cite{Duet_model} uses semantic matching as well as syntactic matching in character $n$-graph  interactions between two documents.
    \item \textbf{MV-LSTM}~\cite{MV_LSTM} is based on a learning goal similar to M-Match. It uses term-term semantic interactions encoded by BiLSTM with positional information.
\end{itemize}

MatchZoo provides the implementations of Dense, Duet, and MV-LSTM. We refer to MatchZoo documentations for their implementation details and settings. We implement BiLSTM using MatchZoo platform. We use a softmax activation function and  cross entropy as the loss function. All models truncate a document to its first 300 tokens. We set epoch to 30 and select the  best model on the dev dataset in terms of AP and report its performance on the test dataset. When training the neural models we use global word embeddings\textsuperscript{\ref{fn:Global_w2v}} instead of local embeddings. Note that our local embeddings trained for each SR do not share embedding spaces.

\begin{figure}[t] \includegraphics[width=0.65\linewidth]{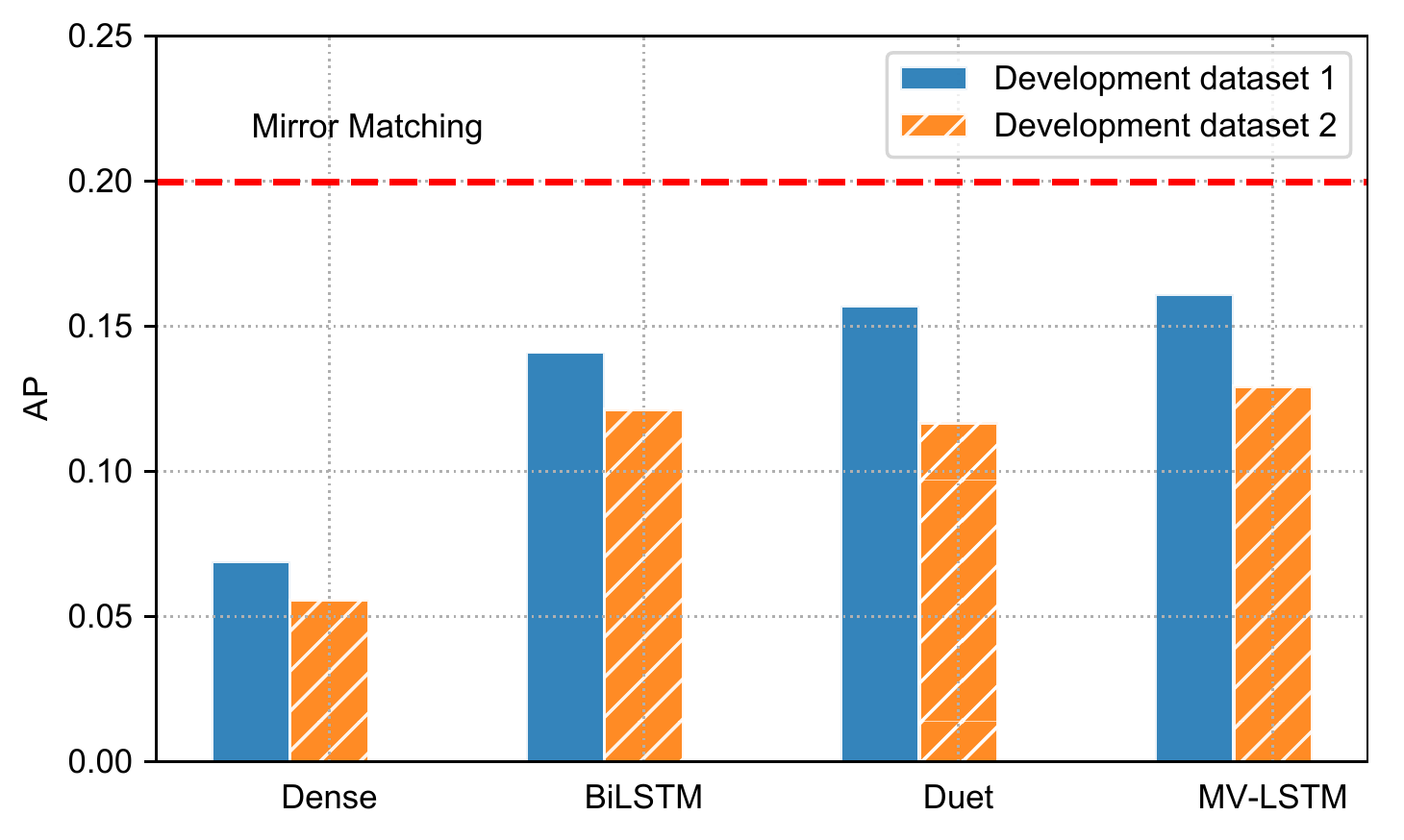}%
\caption{Performance of neural models on test dataset in CLEF2019. Two sets of  AP are presented when the train and dev datasets are comprised of different SRs.}
\label{fig:NeuralIR_Models}
\end{figure}

\paratitle{Results and Discussion.}  Figure~\ref{fig:NeuralIR_Models} presents two sets of  AP when the train and dev datasets are comprised of different SRs. We also show the performance of \mmatch on the test dataset for reference. Here, \mmatch uses the same global word embeddings as the neural models do.
Among all four models, MV-LSTM achieves the highest performance in both evaluation sets, while it is still lower than that of Mirror Matching.

One concern we anticipate is that neural models in ad hoc retrieval are often adopted as a re-ranker. First, documents are ranked by a simple model like BM25 and a neural model then re-ranks the top 1,000 documents, for example. We claim that the setting in screening prioritization in fact similar to the re-ranking setup. Candidate documents in SRs are filtered documents from digital libraries by Boolean retrieval.

It is worth noting that the randomly selected development dataset significantly influences the performance on the test set. One development dataset consistently leads to the higher performance than the other dataset in all models. Moreover, the performance gap between the two development datasets is not negligible.

We believe that different relevance conditions in SRs (or uniqueness of SRs) are a major contributing factor to the performance difference when the train and dev datasets have different splits of SRs. 
Relevant documents within one SR can have higher (in narrow scope of relevance conditions) or lower (in broad scope) inter-similarities than other SRs.
The train/dev instances represent different characteristics when they are comprised of different SRs.
Hence, the patterns learned and validated by one set of train and dev datasets can differ from those by the other set of train and dev datasets. Furthermore, they can be different from patterns required for correct predictions on test SRs.

Another contributing factor can be the size of dataset. Due to the nature of deep neural models, a large amount of data is prerequisite for high performance. It is possible that CLEF2019 dataset might not be sufficient enough to encode all patterns in SRs. More data might mitigate this issue, but it requires further investigations.

\paratitle{Summary.} We examine neural IR models for screening prioritization. The empirical results show large performance changes by different data splits. We discuss a few factors that might be related. To this end, it is hard to draw a conclusion on performance of neural models and how to obtain reliable results remains an open question.

\section{Conclusion}
\label{sec:Conclusion}
We investigate screening prioritization using a query document in the context of document matching. Since the query document is a known relevant document and used for finding unknown relevant documents, we introduce two-way document matching named Mirror Matching. Our proposed model aims to find out how similar two documents (\ie a query document and a candidate document) are. There are two important components in the model that make it  different from alternative approaches, and effective for this task: (1) Two one-way (asymmetric) scores better model to find the most similar two documents. Moreover, it demonstrates that a candidate document can be used as a query if it is a relevant document. (2) Each of the two one-way scores is position-based and it enables PICO element matching as well as semantic matching. This design choice is motivated by our analysis on the distributional patterns of PICO elements in medical literature. Extensive experimental results show the effectiveness of our proposed model compared to unsupervised retrieval models, document similarity measures, and neural models. For future work, we plan to explore approaches for multiple known documents within the document matching framework of screening prioritization.

\bibliographystyle{ACM-Reference-Format}
\bibliography{references}

\end{document}